\documentstyle[aps,twocolumn]{revtex}

\begin{document}
\title{Minority game with arbitrary cutoffs}
\author{N.F. Johnson$^{a}$, P.M. Hui$^{b}$, Dafang Zheng$^{c}$,
and C.W. Tai$^{b,*}$}
\address{$^{a}$ Physics Department, Oxford University\\
Clarendon Laboratory, Oxford OX1 3PU, United Kingdom}
\address{$^{b}$ Department of Physics, The Chinese University of
Hong Kong\\ Shatin, New Territories, Hong Kong}
\address{$^{c}$ Department of Applied Physics,  South China
University of Technology\\ Guangzhou 510641, P.R. China}
\maketitle
\begin{abstract}  We study a model of a competing population of
$N$ adaptive agents, with similar  capabilities, repeatedly
deciding whether to attend a bar with an arbitrary  cutoff $L$. 
Decisions are based upon past outcomes.   The agents are only
told whether the actual attendance is above or below $L$. For
$L \sim N/2$, the game reproduces the main features of Challet
and Zhang's minority game.  As
$L$ is lowered, however, the mean attendances in different runs
tend to divide into two groups.  The corresponding standard
deviations for these two groups are very different.  This 
grouping effect results from the dynamical feedback governing
the game's time-evolution, and is not reproduced if the agents
are fed a random history. 

\vspace*{0.1 true in}



\vspace*{0.1 true in}
\noindent $^{*}$ {\small Present address:  Department of
Applied Physics, The Hong Kong Polytechnic University, Kowloon,
Hong Kong.} 
\end{abstract}

\section{Introduction}

Complex adaptive systems
\cite{holland} have been the subject of much recent attention. 
These systems typically  exhibit rich  global behaviour which
cannot be straightforwardly deduced from the microscopic details
of the constituent objects (agents).  Fascinating 
self-organized phenomena
\cite{bak} have been observed or discussed for a wide range of
systems such as sandpiles, traffic flow, financial markets and
other social phenomena.  Within the physics community,
moreover,  many `microscopic' economics-based models have been
proposed 
\cite{bak1,marsili,amaral} in the growing subfield of
`econophysics'.  

Challet and Zhang \cite{challet} recently introduced a simple
minority game  in which agents repeatedly compete to be in the 
minority group.  These agents have similar capabilities and
each  of them makes decisions based on the past history of
outcomes. Further work on the basic minority game is presented
in Refs.
\cite{savit,johnson2,deCara,rodgers,cavagna}. While the minority
game is stated in terms of agents  choosing between two rooms,
the model can be stated in different ways  to fit different
situations; for example, the agents may as well be deciding
whether to buy or sell a certain stock in a simple market model.
A more  general version of the minority game, namely the
bar-attendance model 
\cite{arthur} in which agents decide whether to attend a bar
with a  certain seating capacity, has also  been studied
recently \cite{johnson1}.  

In this paper, we generalize the minority game to the case of 
arbitrary cutoff.  There are $N$ agents, each  of whom possesses
$s$ strategies, deciding whether to attend a bar with a  seating
capacity, i.e. cutoff, equal to
$L$. Good (bad) decisions correspond  to attending an
undercrowded (overcrowded) bar, or not attending  an overcrowded
(undercrowded) room.  The only information given  to the agents
after each turn is whether the actual attendance was  above or
below the cutoff. We present extensive numerical results for the
mean attendances  and corresponding standard deviations as  a
function of the agents' capabilities and the cutoff
$L$.  For $L=N/2$,  the game reduces to the minority game
\cite{challet,savit,johnson2,deCara,rodgers,cavagna}.  The 
present model for $L\neq N/2$ thus represents an intermediate
model between the minority  game
\cite{challet,savit,johnson2,deCara,rodgers,cavagna} and the
bar-attendance model
\cite{arthur,johnson1}: it allows for arbitrary cutoff  as in
the bar-attendance model, but the actual attendances are not 
announced as in the minority game.  As $L$ is reduced below
$N/2$,  the mean attendances and standard  deviations tend to
be distributed into two groups of values for different  runs of
the same game.  One group comprises a mean attendance which is
insensitive to the agents' capabilities, and is  accompanied 
by a small standard deviation of attendance.  The other group
contains a spread in mean attendances, each with a larger
corresponding standard deviation.  An explanation of these
features is given.

The plan of the paper is as follows.  In Sec. II, the game with 
arbitrary cutoff is defined.  Results for the mean attendances
and  standard deviations for different cutoff $L$, and for
different number  of strategies $s$ per agent, are presented and
discussed in Sec. III.   The relationship  between games with
cutoff $L$ and
$L'=N-L$ is also discussed. Section IV summarizes our main
findings.

\section{The generalized minority game}

Consider a game with $N$ agents deciding whether to go to a bar
with a seating capacity of $L$.  Let the actual attendance at
the bar in the $n$-th turn be $A_{n}$.  If $A_{n} \leq L$ , the
outcome, which is the only information made known to all agents,
is the signal  `undercrowded'. In contrast, if
$A_{n} > L$ then the outcome is the signal `overcrowded'. 
Hence, the  outcome can be represented by a string of zeros
(representing, say, `undercrowded') and ones  (representing
`overcrowded'). The value of the cutoff $L$ is {\em not}
announced.  The agents are not allowed to  communicate among
themselves.  They interact with each other through the common
knowledge of the past history of outcomes, and the fact that
each agent's decision is influencing the others' chances of
being right.  All agents are assumed to have the same level of
capability and to decide the next move based on the most recent
$m$ outcomes. The strategy space, therefore, consists of a
total of
$2^{2^m}$ strategies, analogous to the minority game
\cite{challet}. At the beginning of the game, each agent
randomly picks $s$ strategies from the pool of strategies, with
repetitions allowed during picking.  Each agent uses his best
strategy in making the next decision, i.e. he uses the one with
the highest accumulated merit points.  The merit points are
assigned in the following way.  After the outcome in a turn is
announced, every agent assigns a point to each of his
strategies which would have made the correct decision.  The
correct decisions are attending (not attending) the room with
the outcome being `undercrowded'  (`overcrowded'). The
incorrect decisions are attending (not attending) the room with
the outcome being `overcrowded' (`undercrowded').   The
original minority game \cite{challet} thus  corresponds to
having an odd value of
$N$, and setting $L=N/2$.  We emphasize that our generalized
model is different from the El Farol bar-attendance model
\cite{johnson1} in that the {\em only} publicly-available
information is whether the actual attendance is above or below
the cutoff, i.e. the outcome is a binary digit instead of the
actual attendance. This generalized model may be relevant to
some situations arising in  trading in real  markets.  For
example, when many agents are buying the same  stock, the price
will go up and the majority can then benefit.  In  this case,
the relevant cutoff will be larger than
$N/2$. Alternatively, $L$ may represent some underlying 
`fundamental' value of a particular market.

\section{Results and Discussion}

We have performed numerical simulations for a range of values
of the cutoff $L$.  The number of agents is fixed at
$N=101$.  One of the interesting results in the minority
game is that the mean standard deviation (SD)  or the mean
volatility, i.e. the average of SD's over different runs, shows
a minimum as a function of $m$ for small values of
$s$ (e.g., $s=2,3, \cdots$) \cite{challet,savit}.   The minimum
value of the mean SD is smaller than for the `random' game in
which agents independently decide by tossing a coin
\cite{challet,savit}: the system has thereby managed to
self-organize itself into a state which maximizes the number of
satisfied agents in the community.  This minimum has been
explained by invoking the idea that the effects of the crowd of
agents using the best strategy can become counter-balanced by
the effects of the anticrowd of agents using the corresponding
anti-correlated strategy
\cite{johnson2,rodgers}. 

Figure 1 shows  the standard deviation (SD) and the mean
attendance (inset)  for
$s=2$ as a function of
$m$, with cutoff $L=48$; this represents a small deviation from
the minority game ($L=50$).  For each value of
$m$, 32 runs were carried out using different random initial
conditions.  Each run corresponds to a total of 10000 turns. 
It is clear that there still exists a minimum in the mean SD at
around $m=5$, a feature analogous to the minority game result
\cite{challet,savit}.  For small
$m$, the SD is large since the strategy space is small.  For
large $m$, the mean SD approaches the random coin-toss limit of
$\sigma = \sqrt{N}/2$.  For small $m$, the spread in the mean
attendance is larger.   For large $m$, the mean attendance is
slightly larger than the cutoff and the spread in the mean
attendance is small.  It should be noted that the averaged mean
attendance for $L=48$ is lower than that in the minority game
with $L=50$, indicating the adaptation of the population to the
lower cutoff value: this occurs {\em despite} the fact that the
cutoff $L$ is unknown to the agents. 

Figure 2 shows the SD and mean attendance for $s=2$ and cutoff
$L=40$.  In general, the SD's in the 32 runs spread over a much
larger range of values than in the case with $L=48$.  For the
minority game, the SD's for
$m=2, 3$ are all above the value corresponding to the random
coin-toss limit.  For $L=40$, however,  the SD's at $m=3$ take 
on values {\em below} the coin-toss limit.  An interesting
feature in Fig. 2 is  the tendency of having small SD's in some
runs for all values of
$m$.  A minimum in the averaged SD exists at a smaller value of
$m$ (i.e. $m\approx 3$) than for $L=48$. The values of the mean
attendance at each value of
$m$ also tend to spread over a larger range than for $L=48$. 
The averaged mean attendance is lower than for $L=48$,
demonstrating once more the adaptation of the population to the
(unknown)  cutoff.

As the cutoff $L$ is further reduced, a new feature arises: the
results for the SD and  mean attendance  show the formation of
two separate groups.   One group consists of large SD's with
magnitudes  similar to the minority game, while the other group
consists of  small SD's.   Figure 3 shows the results for
$L=30$.  For each value of $m$,  the upper SD branch is broad
while the lower branch is  narrow.  The lower branch has a
small mean SD (SD
$<1$) and  does not vary much with $m$.  If we take an average
over the two  branches to obtain a mean SD as in the minority
game, the result is  that the minimum in the mean SD gradually
shifts to lower values of 
$m$ as the cutoff is lowered.  However, such an average  may be
misleading as the two branches actually correspond to two
different types  of outcomes (i.e. trajectories) of the game. 
Taken together with the previous results for 
$L=48$ and $40$, the gradual splitting into two branches of
both the  SD and mean attendance distributions is clearly shown
as $L$ decreases.  From this perspective, the minority game with
$L=N/2$ corresponds  to the special case in which the two
branches merge together to form a  rather symmetrical
distribution.  

The formation of a branch with a steady mean attendance together
with a corresponding small SD, for small values of $s$, can be
understood qualitatively as follows.  For small $s$, the chance
for an agent to have picked $s$ strategies with the {\em same}
response to a particular $m$-bit history string, is not
negligible.  Suppose that the recent string of outcomes is
$\cdots 111111$, representing a series of  `overcrowded'
attendances.  Let $N'$ be the number of agents who {\em only}
have strategies with the decision to attend the bar when faced
with an $m$-bit history consisting of just 1's.  If
$N'>L$, these agents will lose, but they will keep attending
since they have no alternative strategies.  The agents who
decide not to attend, i.e. those having strategies with the
decision to stay given the $m$-bit history of 1's, will stay
away and win.  The mean attendance is hence approximately
$N'$ and the corresponding SD is small.  The series of outcomes
$\cdots 111111111\cdots$ thus corresponds to an attractor of the
dynamics for
$N'>L$.

If $N'<L$, the $N'$ agents will attend when faced with an
$m$-bit history of 1's.  Depending on the decisions of the other
$(N-N')$ agents, the next outcome can either be `overcrowded' or
`undercrowded', with a higher chance of overcrowding for $N'
\stackrel{\textstyle <}{\sim} L$.    Therefore, the series
$\cdots 111111\cdots$ with all 1's is no longer an attractor:
the series of outcomes now consists of 0's and 1's.  The mean
attendance will be larger in this case with a larger spread. 
The 0's and 1's in the time series lead to a larger mean SD as
the outcomes make the game look more like the minority game. 
For $N'
\stackrel {\textstyle <} {\sim} L$, the mean attendance will be
higher than the cutoff as there will be more `overcrowded'
attendances than `undercrowded' attendances.  For fixed $N$, the
splitting into two branches appears in the regime of
intermediate
$L$ where the situations $N'<L$ and $N'>L$ occur with comparable
probabilities.  For $s=2$, there will be about $N/4$ agents
picking two strategies with the same response to a given history
on the average.  However, in each run there will be
fluctuations around this mean number.  The game with $L=30$ and
$N=101$ represents the situation in which 
$N'<L$ and $N'>L$ may arise for different runs, and  hence leads
to the formation of the two branches in the mean attendance and
SD. If we further reduce the cutoff, more runs corresponding to
$N'>L$ (i.e. the lower branch) occur.

It has recently been argued that the real history time-series,
and hence the memory, is irrelevant in the minority game
\cite{cavagna}. If the real
$m$-bit history fed to the agents is replaced by a random
$m$-bit string, then the standard deviation of the attendance at
one of the rooms, say 0, is essentially unchanged
\cite{cavagna}. However, replacing the real history by a random
one in the present system does {\em not} reproduce the results
in Fig. 3 for $L=30$.  Both the spreads and values for the {\em
upper}-branch of the SD and mean attendance differ between  the
random and real-history versions. More importantly, the {\em
lower} branch is completely absent in the random-history
version. This is because the  lower branch depends crucially
for its existence on non-local time correlations in the real
history time-series (e.g. $\cdots 11111\cdots$). Our findings
therefore demonstrate the possible dangers of replacing the true
history by a random one when evaluating dynamical properties of 
such complex adaptive systems. We note in passing that even in
the basic minority game with
$L=N/2$, correlations do arise in the real history time-series
\cite{savit,johnson3}.

Figure 4 shows the mean attendance for $L=70$ and $s=2$.  From
the definition of the game, games with cutoff $L' \equiv N-L$
are related to games with cutoff $L$ in that the mean
attendance in the game with
$L'$ can be found from the mean population of agents {\em not}
attending the room in a game with $L$. It is, therefore,
sufficient to study the range of values with $L \leq N/2$.
Comparing the results in Fig. 3 ($L = 30$)  with those in Fig.
4 ($L'= 70$), the symmetry between games with $L'$ and
$L$ is clearly shown.

Figures 5 and 6 show results with a larger value of $s$ ($s=5$)
for the cases of $L=40$ and $30$, respectively.  For $L=40$,
the  minimum in the mean SD can clearly be seen at
$m=5$.  For $L=30$ the  minimum, although not as sharp as for
larger values of $L$, is at $m=4$: for larger $m$, the SD's
start to spread out and show the tendency to  distribute into
two groups.   Comparing the results at cutoff $L=30$ for
$s=2$ (Fig. 3) and  
$s=5$ (Fig. 6), the formation of the two branches is much more 
pronounced for small 
$s$.  This is consistent with our argument above since the
chance that an agent picks $s$ strategies with the same
response to  a particular $m$-bit history is smaller for larger
values of
$s$.  Hence, having $L=30$ and $s=5$ corresponds to the case
with 
$N'\ll L$: the resulting behavior is hence similar to a minority
game, with the exception that some runs have smaller SD's.  It
should also be  noted from the results for both $s=2$ and
$s=5$ that the minimum in  the mean SD gradually shifts to a
lower value of $m$ as the cutoff is  reduced.  

\section{Conclusions}

A generalized version of the minority game with arbitrary 
cutoff $L$ was proposed and studied.  Features in the mean
attendance and standard deviation can be quite different from
the basic minority game as the cutoff shifts  away from
$N/2$.  In particular, the mean attendances in different runs
tend to divide into two groups.  The corresponding standard
deviations for these two groups are very different.  These
features are not reproduced if the agents are fed a random
history, thereby demonstrating the importance of dynamical
feedback and hence memory in this system.

\begin{center} {\bf ACKNOWLEDGMENTS}
\end{center}

We thank D. Challet, D. Leonard, D. Sherrington and A. Cavagna
for discussions concerning the basic minority game. On of us
(D.Z.) would like to thank the Department  of Physics at the
Chinese University of Hong Kong for partial  support through a
C.N. Yang Visiting Fellowship.

\centerline{\bf Figure Captions}

\bigskip

\noindent Figure 1: The standard deviation (SD) and mean
attendance (inset)  as a function of the memory $m$ for $N=101$
agents,
$s=2$ strategies per agent and bar cutoff
$L=48$.  For  each value of $m$, data from 32 different runs are
shown.  

\bigskip

\noindent Figure 2: Same as in Figure 1 for $N=101$, $s=2$ and
$L=40$. 

\bigskip

\noindent Figure 3: Same as in Figure 1 for $N=101$, $s=2$ and
$L=30$. 

\bigskip

\noindent Figure 4: Same as in Figure 1 for $N=101$, $s=2$ and
$L=70$.   Note the relationship between the games with $L=30$
(Figure 3) and
$L=70$. 

\bigskip

\noindent Figure 5: The standard deviation and mean attendance
(inset) as  a function of $m$ for $N=101$, $s=5$ and cutoff
$L=40$.  For each value  of $m$, data from 32 different runs are
shown. 

\bigskip

\noindent Figure 6: Same as in Figure 5 for $N=101$, $s=5$ and
$L=30$. 

\end{document}